\documentclass[preprint2,times]{aastex62}
\usepackage{amsmath}

\shorttitle{sloshing oscillations in hot loops}
\shortauthors{Krishna Prasad et al.}

\begin{document}
\title{Compressive oscillations in hot coronal loops: Are sloshing oscillations and standing slow waves independent?}

\correspondingauthor{S. Krishna Prasad}
\email{krishna.prasad@kuleuven.be}

\author[0000-0002-0735-4501]{S. Krishna Prasad} 
\affiliation{Centre for mathematical Plasma Astrophysics, Department of Mathematics, KU Leuven, Celestijnenlaan 200B, B-3001 Leuven, Belgium}                              
\author{T. Van Doorsselaere}
\affiliation{Centre for mathematical Plasma Astrophysics, Department of Mathematics, KU Leuven, Celestijnenlaan 200B, B-3001 Leuven, Belgium}  

\begin{abstract}
Employing high-resolution EUV imaging observations from SDO/AIA, we analyse a compressive plasma oscillation in a hot coronal loop triggered by a C-class flare near one of its foot points as first studied by Kumar et al. We investigate the oscillation properties in both the 131{\,}{\AA} and 94{\,}{\AA} channels and find that what appears as a pure sloshing oscillation in the 131{\,}{\AA} channel actually transforms into a standing wave in the 94{\,}{\AA} channel at a later time. This is the first clear evidence of such transformation confirming the results of a recent numerical study which suggests that these two oscillations are not independent phenomena. We introduce a new analytical expression to properly fit the sloshing phase of an oscillation and extract the oscillation properties. For the AIA 131{\,}{\AA} channel, the obtained oscillation period and damping time are 608$\pm$4{\,}s and 431$\pm$20{\,}s, respectively during the sloshing phase. The corresponding values for the AIA 94{\,}{\AA} channel are 617$\pm$3{\,}s and 828$\pm$50{\,}s. During the standing phase that is observed only in the AIA 94{\,}{\AA} channel, the oscillation period and damping time have increased to 791$\pm$5{\,}s and 1598$\pm$138{\,}s, respectively. The plasma temperature obtained from the DEM analysis indicates substantial cooling of the plasma during the oscillation. Considering this, we show that the observed oscillation properties and the associated changes are compatible with damping due to thermal conduction. We further demonstrate that the absence of a standing phase in the 131{\,}{\AA} channel is a consequence of cooling plasma besides the faster decay of oscillation in this channel.
\end{abstract}


\section{Introduction}
Flare associated hot loops, with plasma temperatures exceeding 6 MK, often display compressive oscillations. These oscillations were first discovered in Doppler velocities of hot spectral lines such as Fe{\,}\textsc{xix} and Fe{\,}\textsc{xxi} observed by the Solar Ultraviolet Measurement of Emitted Radiation (SUMER) instrument onboard the \textit{Solar and Heliospheric Observatory} (SoHO) \citep{2002ApJ...574L.101W}. Their oscillation period was in the range of 7--31 minutes and their phase speed was estimated to be close to the local acoustic speed \citep{2003A&A...406.1105W}. Additionally, in some cases, a quarter period phase difference is found between the Doppler velocity and the corresponding intensity oscillations leading to the interpretation of these oscillations as standing slow magneto-acoustic waves \citep{2003A&A...402L..17W}. Another characteristic feature of these oscillations is that they exhibit rapid damping with decay times on the same order as the oscillation period. Similar oscillations were observed by the Bragg Crystal Spectrometer (BCS) onboard \textit{Yohkoh} \citep{2005ApJ...620L..67M, 2006ApJ...639..484M} and the Extreme ultraviolet Imaging Spectrometer (EIS) onboard \textit{Hinode} \citep{2008ApJ...681L..41M}. However, the period of oscillations measured from the BCS data is on the lower end with an average value of 5.5$\pm$2.7 minutes \citep{2006ApJ...639..484M}. This discrepancy is explained in terms of the possible observation of shorter loops by BCS as it is sensitive to much hotter plasma ($\approx$12 MK). Since these oscillations are usually preceded by a brightening near one of the foot points, it is believed that a micro-flare or similar reconnection event triggers them \citep{2003A&A...406.1105W, 2005A&A...435..753W}. Thermal conduction has been shown to be a major cause for their damping \citep{2002ApJ...580L..85O} although recent studies indicate compressive viscosity \citep{2015ApJ...811L..13W, 2018ApJ...860..107W} or thermal misbalance \citep{2017ApJ...849...62N, 2019A&A...628A.133K} can dominate depending on the physical conditions within the loop. A number of theoretical and numerical investigations were also made including the application of forward modelling techniques in some cases, especially to study their driving mechanism \citep{2005A&A...436..701S, 2009AnGeo..27.3899S, 2012ApJ...754..111O,2018ApJ...860..107W}, their damping behaviour \citep{2008A&A...483..301B, 2008ApJ...685.1286V, 2013A&A...553A..23R, 2014ApJ...786...36A, 2018ApJ...860..107W} and to predict/reproduce some of their observational characteristics \citep{2004A&A...414L..25N, 2006A&A...446.1151N, 2015ApJ...807...98Y, 2015ApJ...813...33F}. We refer the interested reader to comprehensive reviews by \citet{2011SSRv..158..397W} and \citet{2021SSRv..217...34W} on this subject.

Compressive oscillations are observed in other coronal structures too \citep{2006RSPTA.364..473N, 2009SSRv..149...65D, 2011SSRv..158..267B}. Indeed, they are found to be ubiquitous in open/extended loop structures which are relatively quiescent and cold as compared to the hot flare loops \citep{2012A&A...546A..50K, 2018ApJ...853..145M}. However, these oscillations are mainly due to driven waves and are believed to be originated in the lower solar atmosphere \citep{2011ApJ...728...84B, 2012ApJ...746..119R, 2012ApJ...757..160J, 2015ApJ...812L..15K}. Their oscillation periods also range from few minutes to few tens of minutes and they exhibit rapid damping similar to that of standing waves \citep{2006A&A...448..763M}. There have been extensive studies on their characteristic properties, especially on their damping behaviour in the solar corona, by a number of authors \citep[][to name but a few]{2003A&A...408..755D, 2004A&A...415..705D, 2009SSRv..149...65D, 2012A&A...546A..50K, 2014ApJ...789..118K, 2016GMS...216..419B, 2019FrASS...6...57S}. Please refer to \citet{2020arXiv201208802B} for a recent review on these waves.

Analysing the microwave emission from a hot flare loop, \citet{2012ApJ...756L..36K} have shown that the associated plasma density exhibits rapidly decaying oscillations with a periodicity of 12.6 minutes and a decay time of about 15 minutes. In addition, the co-temporal high-resolution imaging observations of the loop, acquired by the Atmospheric Imaging Assembly (AIA) onboard the \textit{Solar Dynamics Observatory} (SDO), also display similar oscillations. Based on the observed characteristics, the authors interpreted these oscillations as due to the standing slow magneto-acoustic waves. Subsequently, the high-resolution imaging data from \textit{SDO}/AIA have revealed spatially resolved longitudinal oscillations in hot coronal loops \citep{2013ApJ...779L...7K, 2015ApJ...804....4K}. These oscillations involve a plasma perturbation bouncing back and forth between the two foot points of the loop while its amplitude decays rapidly. They are interpreted as reflected propagating slow waves. Later studies refer to them as `sloshing' oscillations \citep{2016ApJ...826L..20R, 2019ApJ...874L...1N}. The general properties of these oscillations are very similar to that observed by SUMER. Furthermore, these oscillations too appear to be driven by a small flare near one of the foot points. However, as the spatiotemporal properties of these oscillations do not resemble a standing wave, their association with the standing slow waves discovered by SUMER is not very clear. \citet{2015ApJ...811L..13W} report a unique event from the SDO/AIA observations of hot coronal loops, where the authors find a clear evidence for standing slow waves. The general properties are again very similar but, in this case, the oscillation does not display significant spatial movements rather it exhibits anti-phase perturbations between the two legs resembling a standing mode. Additionally, employing 1D nonlinear MHD simulations \citet{2018ApJ...860..107W} have shown that a pressure disturbance generated by a flare-like impulsive event would initially bounce back and forth between the foot points before transforming into a standing slow wave. They further  demonstrated that this transformation can occur immediately after the first reflection if the physical conditions within the loop are favourable to efficiently dissipate higher harmonics. Thus, the authors were able to explain both the sloshing oscillations and the standing slow waves with a common interpretation. However, so far from observations we have only seen either a sloshing oscillation \citep{2013ApJ...779L...7K, 2015ApJ...804....4K, 2016ApJ...828...72M, 2017A&A...600A..37N} or a standing wave \citep{2015ApJ...811L..13W} but there has not been ample evidence to support the transformation between them. This raises an important question whether the sloshing oscillations are an independent class of oscillations \citep{2019ApJ...874L...1N}.

In this letter, we analyse the same event studied by \citet{2013ApJ...779L...7K} which was interpreted by them as a reflected propagating (or sloshing) wave. We reveal a number of interesting properties including a clear evidence for the transformation of sloshing oscillation into a standing wave. In Section \ref{obs} we present the details on observations, followed by our analysis and results in Section \ref{anres}, and finally list our conclusions in Section \ref{concl}.

\section{Observations}
\label{obs}
The dataset employed in this study is the same as that analysed by \citet{2013ApJ...779L...7K} except that here we extended it to a longer duration. The data mainly constitute imaging observations of a hot coronal loop performed by the \textit{SDO}/AIA \citep{2012SoPh..275...17L, 2012SoPh..275....3P}. A subfield of about 280\arcsec$\times$280\arcsec\ encompassing the target loop structure within the active region NOAA 11476, is considered (see Fig.{\,}\ref{tdmaps}). The image sequences obtained from 17:20 UT until 18:46 UT on May 7, 2012, across 6 wavelength channels, namely the AIA 94{\,}{\AA}, 131{\,}{\AA}, 171{\,}{\AA}, 193{\,}{\AA}, 211{\,}{\AA}, and 335{\,}{\AA} channels, were analysed. All the data were processed to level 1.5, incorporating the necessary instrumental corrections by following standard procedures. In particular, the roll angle and plate scale corrections and the co-alignment across multiple wavelength channels were achieved by using a robust pipeline that is publicly available\footnote{http://www.staff.science.uu.nl/~rutte101/rridl/sdolib/}. The final pixel scale and the cadence of the data are 0$\farcs$6 and 12{\,}s, respectively.

\section{Analysis and Results}
\label{anres}
\subsection{Time-distance maps}
\begin{figure*}
\centering
\includegraphics[scale=0.5]{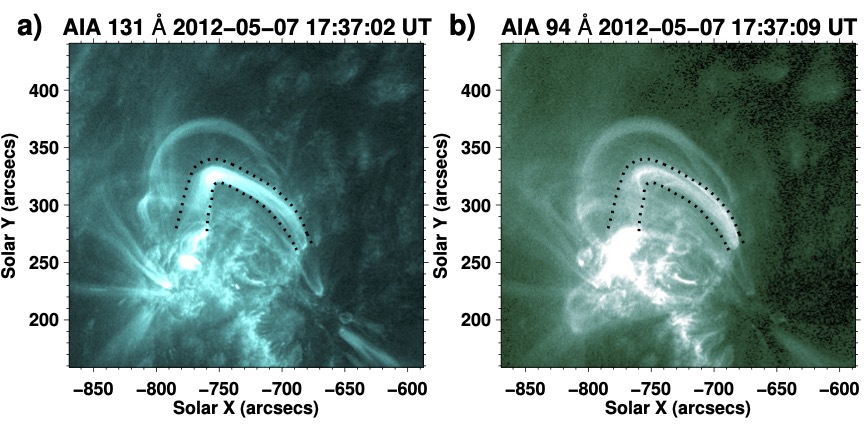}
\includegraphics[scale=0.52]{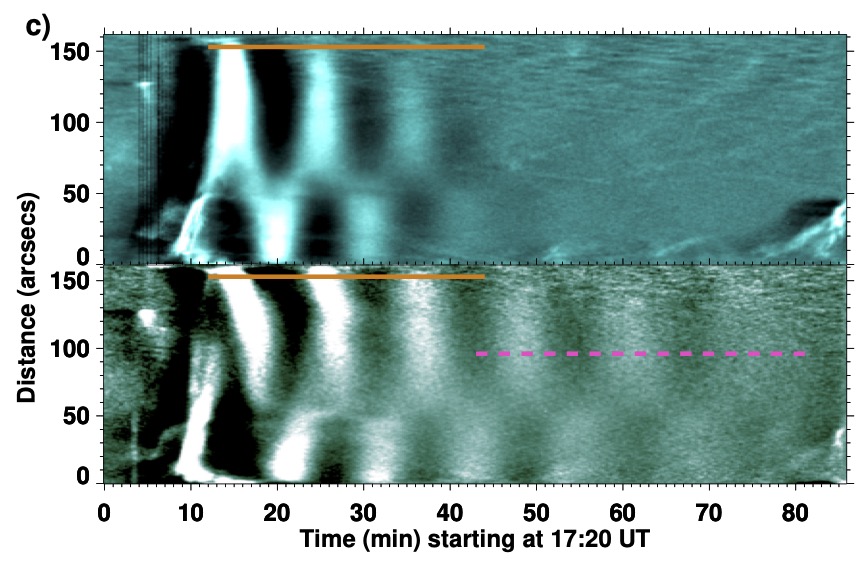} 
\caption{a) A snapshot of the target loop structure and its vicinity within NOAA AR11476 as observed from the AIA 131{\,}{\AA} channel. The black dotted lines mark the boundaries of the selected region encompassing the loop. b) Same as panel a but for the AIA 94{\,}{\AA} channel. c) Time-distance maps constructed from the selected loop region by averaging intensities across the loop. The top and bottom plots show the respective maps for the AIA 131{\,}{\AA} and 94{\,}{\AA} channels. The distances on $y$-axes begin from the wider end of the loop. The horizontal pink dashed and orange solid lines marked over these maps represent the locations and the sections of the time series that are selected for the subsequent analyses.}
\label{tdmaps}
\end{figure*}

Snapshots of the analysed loop structure, one each from the AIA 131{\,}{\AA} and 94{\,}{\AA} channels, are shown in Fig.{\,}\ref{tdmaps}. In these wavelength channels, the corresponding image sequences display sloshing oscillation of plasma along the loop following a C-class flare near one of its foot points. In order to study its evolution, we constructed time-distance maps by manually selecting the boundaries of the loop (black dotted lines in Figs.{\,}\ref{tdmaps}a \& \ref{tdmaps}b) and averaging the transversal intensities within that region from the individual images. Subsequently, to enhance the visibility of oscillations, the intensities at each spatial position were detrended and normalised using a twelve-minute (60 point) running average of the corresponding time series. It may be noted that the detrending is done here only to increase the contrast of the oscillation. Besides, the oscillation pattern visible in the original and the detrended time-distance maps is congruent assuring that this process did not introduce any unwanted artefacts. The resultant maps are displayed in panel `c' of the figure with the top and bottom plots corresponding to that from AIA 131{\,}{\AA} and 94{\,}{\AA} channels, respectively.
The vertical dark stripes visible at the beginning of the time series in AIA 131{\,}{\AA} are due to alternating low-exposure frames that are automatically acquired by AIA in specific wavelength channels during flares. These are unavoidable when analysing the data at full cadence but in order to minimise their effect the intensities in individual frames were normalised by the respective exposure times at the beginning. Bright slanted ridges forming a `triangular' wave pattern are apparent in the time-distance maps for both the channels. This suggests back and forth motion of the plasma indicating the presence of sloshing oscillation in the loop. As can be seen, its evolution is not the same across the two channels. In the AIA 131{\,}{\AA} channel, the oscillation decays rapidly and does not seem to possess detectable amplitude beyond three cycles as noted previously by \citet{2013ApJ...779L...7K}. However, in AIA 94{\,}{\AA} channel, the oscillation appears to have appreciable signal for up to six cycles! This differential behaviour, likely suggesting a multi-thermal nature of the loop, has not been reported by previous authors but it is rather interesting. Another striking feature evident in the 94{\,}{\AA} channel, is the eventual transformation of sloshing oscillation (triangular ridges) into a standing wave (vertical ridges). Although, this evolution is in excellent agreement with the previous nonlinear MHD simulations \citep{2012ApJ...754..111O, 2018ApJ...860..107W}, this is the first time we have a clear observational detection of such transformation. It may be noted that the earlier hydrodynamic simulations by \citet{2016ApJ...826L..20R} do not seem to show any transformation in the evolution of sloshing oscillations in a coronal loop. Furthermore, until now there is only separate evidence for either a sloshing oscillation or a standing wave prompting us to think if the sloshing oscillations are a different class of oscillations. Therefore, it is remarkable to see what has been earlier reported to be a sloshing oscillation transform itself into a standing wave, albeit in a different temperature channel, suggesting that these two phenomena are not independent but rather part of the same event. 

\subsection{Properties of oscillations}
As the oscillation appears to slowly transform from a sloshing phase into a standing phase, we derive the properties separately for these two phases as described in the following sections.

\subsubsection{Sloshing phase}
\label{sloshph}
The sloshing phase of the oscillation is visible in both the channels. Although it has been a standard practice to pick the time series from a particular spatial position and fit a damping sine function (as we show for the standing phase of the oscillation in Section{\,}\ref{standph}), here we demonstrate that this is not a correct procedure for sloshing oscillations. Because the perturbation exhibits significant spatial movements during this phase, it is not as trivial to extract the oscillation parameters. We also show that the oscillation parameters are dependent on the spatial location from where the time series is extracted. Furthermore, we provide a better model to properly fit and derive the properties of sloshing oscillations.

In order to simulate the sloshing oscillations, we approximate the initial perturbation as a Gaussian whose amplitude is decaying with time, while its location is oscillating between the two end points of a line segment. This model can be described by the following equation.
\begin{equation}
	I(x,t)=A_{0}{\,}e^{(-t/\tau)}{\,}e^{\left((x-x_{0})^{2}/\sigma^{2}\right)}
	\label{eq1}
\end{equation}
\begin{equation*}
\mathrm{where,}~ x_{0} = \frac{L}{2}\left(1-\mathrm{cos}\left(2\pi\frac{t}{P}+\phi\right)\right).
\end{equation*}
Here, $x$ is the spatial coordinate, $t$ is time, $L$ is the length of the structure, $P$ is the period of the sloshing motion, $\tau$ is the damping time, and $\sigma$ is the width of the perturbation. $A_0$ and $\phi$ are constants. 
\begin{figure*}
\centering
\includegraphics[scale=0.35]{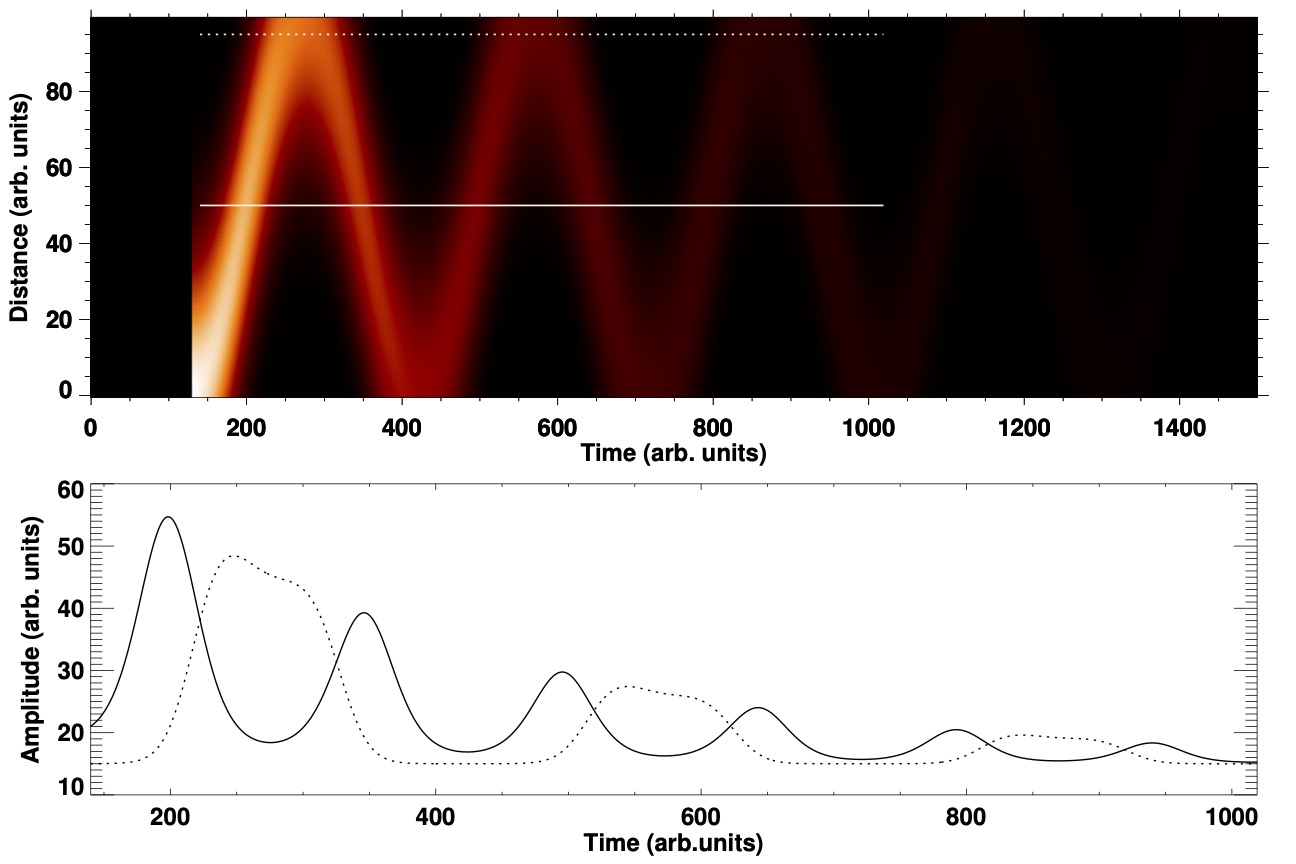} 
\caption{A toy model imitating the sloshing oscillations. The top panel displays the spatio-temporal evolution of a decaying Gaussian bouncing back and forth between the end points of a linear structure (see Equation{\,}\ref{eq1}). A constant background and an offset at the beginning of the oscillation are added to replicate the observations. The solid and dotted lines mark the locations of the extracted light curves that are shown in the bottom panel with the same line style. An animation displaying the temporal evolution of the Gaussian perturbation is available online. The duration of the animation is 10{\,}s.}
\label{tdmodel}
\end{figure*}
Using an arbitrary set of values for all these parameters, a sample sloshing oscillation has been generated whose spatio-temporal evolution is shown in the top panel of Fig.{\,}\ref{tdmodel}. An offset at the beginning of the oscillation and a constant background are added on purpose to mimic the observations. As can be seen, the triangular wave pattern broadly describes the sloshing oscillations observed. The temporal evolution of the oscillation at two spatial locations, one near the edge (dotted line) and another near the centre (solid line), are shown in the bottom panel of the figure. These time series are clearly non-sinusoidal. Indeed, such non-sinusoidal signatures with flat bottoms in the light curves were commonly observed in the previous studies (for e.g., see Fig.{\,}3 of \cite{2013ApJ...779L...7K} and Fig.{\,}4 of \citet{2016ApJ...828...72M}). The time series near the centre does not appear as deviant but one must note that this behaviour is heavily dependent on the width of the perturbation. Moreover, it is evident that the oscillation period near the centre is twice as much as that near the edge. This is not very surprising since at any location away from the edges, the perturbation crosses twice before completing one full oscillation. Therefore, we emphasize that fitting the time series from any spatial location with a damping sinusoidal function does not provide an accurate description of sloshing oscillation properties.

Ideally, one could use the model described by Equation{\,}\ref{eq1} to perform a two dimensional fit with the time-distance maps from observations (similar to \citet{2020ApJ...898..126P}) and extract important oscillation properties. However, because the spatial coordinate in observations is projected and any nonplanar structuring in the loop makes this projection non-uniform along its length, a direct comparison with such models is not possible. To circumvent this problem, we fix the spatial coordinate in our model to one of the end points, i.e., $x=0$, which reduces Equation{\,}\ref{eq1} to 
\begin{equation}
	I(0,t)=A_{0}{\,}e^{(-t/\tau)}{\,}e^{\left(-x_{0}^{2}\right)}
	\label{eq2}
\end{equation}
\begin{equation*}
\mathrm{where,}~ x_{0} = \frac{1}{2 \sigma_n}\left(1-\mathrm{cos}\left(2\pi\frac{t}{P}+\phi\right)\right).
\end{equation*}
Here $\sigma_n = \sigma/L$ is the normalised width of the perturbation. Equation{\,}\ref{eq2} is now independent of the spatial coordinate and therefore can be directly applied to a time series obtained near the foot point of a loop exhibiting sloshing oscillations. To implement this, we first extract a relevant section of the time series (marked by orange solid lines in Fig.{\,}\ref{tdmaps}c) from both the channels. The original intensities are then detrended and normalised using the background generated from fitting a parabolic curve to the minima of the oscillation.
\begin{figure*}
\centering
\includegraphics[scale=0.4]{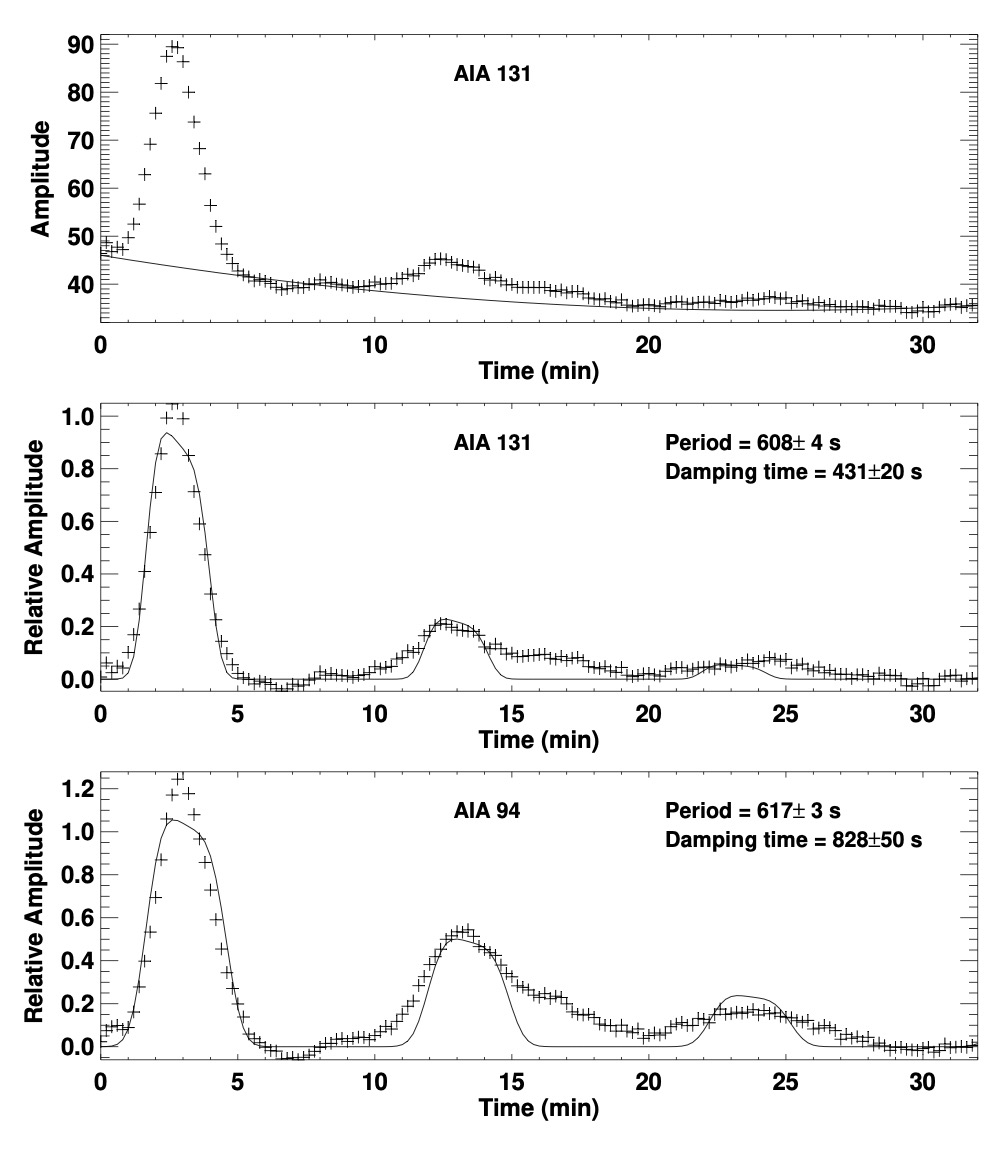} 
\caption{Extraction of oscillation parameters during the sloshing phase. The top panel displays the original light curve from the AIA 131{\,}{\AA} channel extracted from a location marked by the orange solid line in Fig.{\,}\ref{tdmaps}c. The solid line here represents a background constructed by fitting a parabolic curve to the oscillation minima. The middle panel displays the resultant detrended and normalised light curve. The solid line in this plot represents the best fit to the data following Equation.{\,}\ref{eq2}. The oscillation period and damping time obtained from the fitted curve are listed in the plot. The bottom panel shows the corresponding plot for the AIA 94{\,}{\AA} channel.}
\label{rwfit}
\end{figure*}
It may be noted that here we did not use a running average method that is commonly employed for the construction of background. This is because the perturbation does not appear to be a symmetric modulation during the sloshing phase. The original time series and the constructed background from the AIA 131{\,}{\AA} channel are shown in the top panel of Fig.{\,}\ref{rwfit} for illustration. As can be seen, the parabolic curve provides a good approximation for the background. The resultant intensities after the background subtraction and normalisation are presented in the middle and the bottom panels for the AIA 131{\,}{\AA} and 94{\,}{\AA} channels respectively. The best fits to the data obtained by fitting the model in Equation{\,}\ref{eq2} via chi-square minimisation, are shown as solid curves in these panels. The derived oscillation periods from these fits are 608$\pm$4{\,}s and 617$\pm$3{\,}s and the respective damping times are 431$\pm$20{\,}s and 828$\pm$50{\,}s for the oscillations in AIA 131{\,}{\AA} and 94{\,}{\AA} channels.

For comparison, the oscillation period and damping time obtained by \citet{2013ApJ...779L...7K} from the AIA 131{\,}{\AA} channel are 634{\,}s and 437{\,}s, respectively. These values are not very different from those derived here. The reason for this is two fold. Firstly, the authors also preferred a time series closer to the loop foot point despite the one closer to the apex being more sinusoidal since the latter exhibits a double peak. Secondly, the authors average the signal over a large spatial region reducing any substantial deviations from a sine curve. Furthermore, in this example, the spatial extent of the perturbation is quite large (as evident by the merging of the oncoming and forward going part of the perturbation near the loop apex) which is not always the case. We would expect the discrepancies from a simple sine model to be much larger when the spatial extent of the perturbation is small compared to the length of the loop \citep[see for e.g., Fig.{\,}6 of][]{2016ApJ...828...72M}. Hence, as such, our model is widely applicable.

Nevertheless, we note some of the important caveats of our model. The Gaussian approximation to the perturbation may not be always reasonable. We did not take into account the spatial/temporal changes in the width of the perturbation. Also, the transition from a sloshing phase to a standing phase of the oscillation is a gradual process so picking a section of the time series to represent either of the phases is not always obvious. Some of these limitations could be the reason why there is a larger discrepancy between the model and the data during the later part of the oscillation (see Fig.{\,}\ref{rwfit}).

\subsubsection{Standing phase}
\label{standph}
The standing phase of the oscillation is only apparent in the 94{\,}{\AA} channel. In order to extract its properties, we take the appropriate section of the time series at a selected spatial position along the loop (marked by the pink dashed line in Fig.{\,}\ref{tdmaps}c) where the oscillations are clearly discernible.
\begin{figure*}
\centering
\includegraphics[scale=0.45]{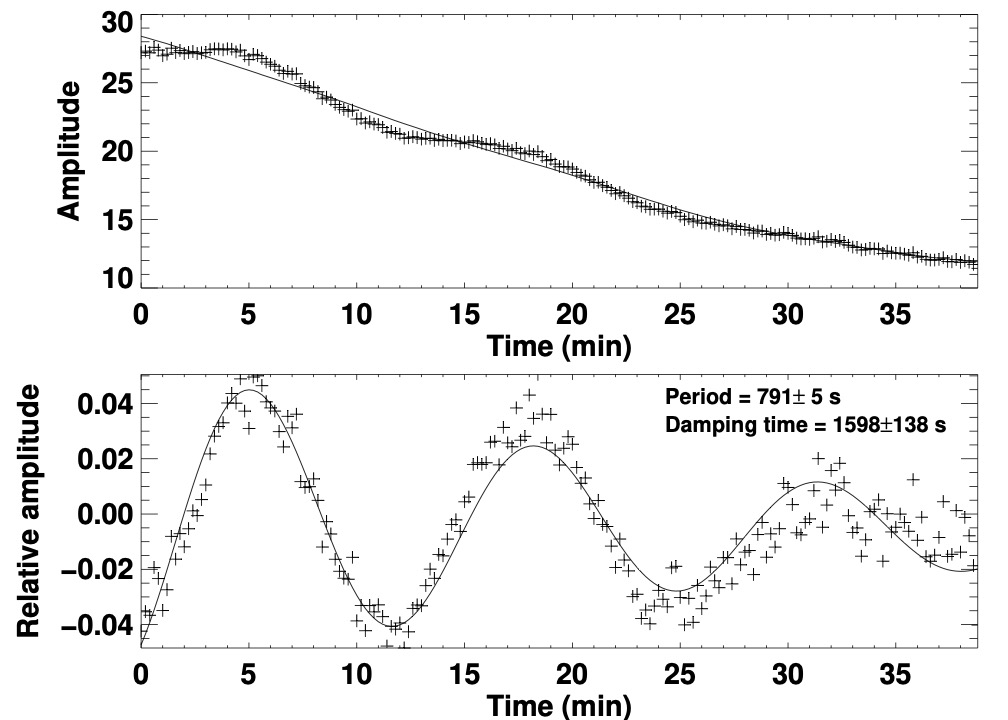} 
\caption{Extraction of oscillation parameters during the standing phase. The top panel displays the original light curve extracted from the location marked by a pink dashed line in Fig.{\,}\ref{tdmaps}c. The solid line in this plot represents a background constructed from the twelve-minute running average of the data. The bottom panel shows the corresponding detrended and normalised light curve. The solid line here represents the best fit damping sine curve following Equation{\,}\ref{eq3}. The oscillation period and damping time obtained from the fitted curve are listed in the plot.}
\label{swfit}
\end{figure*}
The original intensities are first detrended and normalised using a background constructed from a twelve-minute running average of the time series and the resultant intensities are then fit with a simple damping sine function given by the following equation
\begin{equation}
	I(t) = A_{0}{\,}e^{(-t/\tau)}{\,}\mathrm{sin}\left(2\pi\frac{t}{P}+\phi\right) + B_{0} + B_{1}t 
\label{eq3}
\end{equation}
where $t$ is time, $P$ is the oscillation period, and $\tau$ is the damping time. $A_0$, $\phi$, $B_0$, and $B_1$ are constants. The best fit thus obtained (via chi-square minimisation) is shown in the bottom panel of Fig.{\,}\ref{swfit} as a solid line over the normalised data. The original intensities and the constructed background are also shown in the top panel of this figure. The oscillation period and the damping time computed from the fitted curve are 791$\pm$5{\,}s and 1598$\pm$138{\,}s respectively.

\subsection{DEM analysis}
By comparing the oscillation properties between the sloshing phase and the standing phase in the AIA 94{\,}{\AA} channel, it is evident that the damping time has increased nearly by a factor of 2 during the latter phase. The oscillation period has also increased although by a lesser amount. To understand these changes, we seek to find how the plasma thermal properties vary over the duration of the oscillation. 
\begin{figure*}
\centering
\includegraphics[scale=0.47]{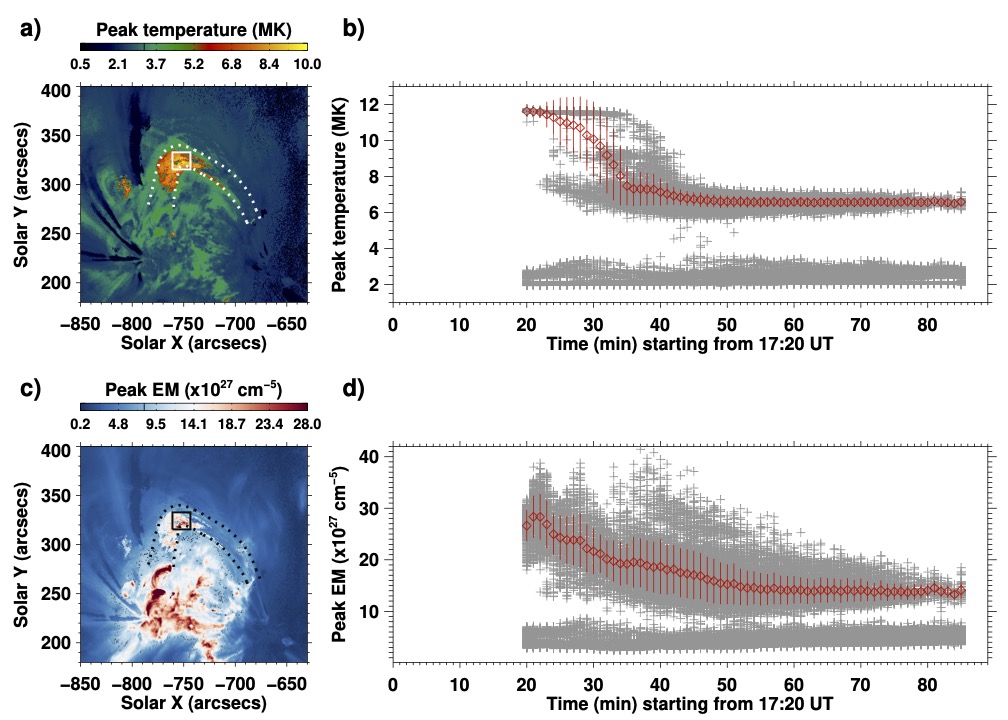} 
\caption{a) A sample peak temperature map obtained from the DEM analysis. The white dotted lines highlight the location of the loop. The white square marks a selected region near the loop apex over which the evolution of the temperature is plotted in panel b. b) Grey plus symbols denote the temperature values obtained from all the pixels within the square region shown in panel a, as a function of time. The red diamond symbols and the vertical bars denote the mean and standard deviation values, respectively, at each instant after excluding the outliers below 4{\,}MK. c) The corresponding peak emission measure map. The locations of the loop and the selected square region near the apex are shown here with black lines. d) Same as panel b but for emission measure. The outliers in this plot are considered as those with emission measure values below 8$\times$10$^{27}${\,}cm$^{-5}$.  }
\label{emte}
\end{figure*}
Employing a regularised inversion code developed by \citet{2012A&A...539A.146H}, we perform Differential Emission Measure (DEM) analysis using the observed intensities in all 6 AIA coronal channels (94{\,}{\AA}, 131{\,}{\AA}, 171{\,}{\AA}, 193{\,}{\AA}, 211{\,}{\AA}, and 335{\,}{\AA}). DEMs were computed at each spatial location but for every fifth frame lowering the cadence of these data to 1 minute to reduce computation time. Also, the calculations were not done for the data within the first 20 minutes i.e., between 17:20 UT and 17:40 UT, as there were many low-exposure frames in some wavelength channels during this period. Subsequently, the temperature ($T$) and emission measure (EM) corresponding to the peak emission in the DEM curves are noted and used to build maps of these parameters at each time step. A sample temperature map and the corresponding emission measure map thus obtained are shown in Fig.{\,}\ref{emte}. In order to study the temporal changes in these quantities, we choose a region near the loop apex (marked by black/white squares in Fig.{\,}\ref{emte}a \& c) and plot the temperature and emission measure values obtained from all the pixels within this region as a function of time in panels `b' and `d' of this figure (grey '+' symbols). Many outliers with substantially lower values of temperature/emission measure as compared to the majority have been detected in these plots. These are due to bad DEM fits arising from low signal locations and possibly represent the background. In any case, since we are interested in the hot plasma we exclude these outliers and compute the average values in the individual parameters at each time step. These values are marked by red diamond symbols in the figure. The corresponding standard deviation values are shown as vertical bars on these symbols. As can be seen from these plots both the temperature and emission measure decrease substantially during the oscillation. A steeper decline in temperature is found until about 17:55 UT after which the decrease is more gradual. In contrast, the emission measure decreases gradually throughout the duration. Nonetheless, the general decrease in these parameters is natural since the dense hot plasma injected into the loop should eventually cool down and get dispersed.

\begin{table*}
\begin{center}
\caption{Comparison of oscillation properties between the sloshing (17:40 UT) and the standing phases (18:20 UT).}
\label{tab1}
\begin{tabular}{c c c c c c c c}
\hline\hline
Time & $T$ (MK) & $\rho$ (cm$^{-3}$) & \multicolumn{2}{c}{$P$ (s)} & \multicolumn{2}{c}{$\tau$ (s)} & $\tau^{\prime}$ (s)\\
     &          &    & AIA 131{\,}\AA  & AIA 94{\,}\AA & AIA 131{\,}\AA  & AIA 94{\,}\AA &             \\
 \hline
17:40 UT & 11.6 & 5.6$\times$$10^9$ & 608$\pm$4 & 617$\pm$3 & 431$\pm$20 & 828$\pm$50 & 945 \\
18:20 UT & 6.6 & 4.1$\times$$10^9$ & \nodata & 791$\pm$5 & \nodata & 1598$\pm$138 & 2396 \\
\hline
\end{tabular}
\end{center}
\tablecomments{$\tau^{\prime}$ is damping time due to thermal conduction estimated from linear wave theory.}
\end{table*}

To assess the impact of these changes in plasma properties on the oscillation, we select two specific instants in time, 17:40 UT and 18:20 UT, as a representative of the sloshing and standing phases of the oscillation, respectively. During the period between these two instants, the temperature dropped from about 11.6 MK to 6.6 MK while the corresponding emission measure decreased from about 2.7$\times$10$^{28}${\,}cm$^{-5}$ to 1.4$\times$10$^{28}${\,}cm$^{-5}$. Assuming the emission depth as equivalent to the width of the loop ($w$) (i.e., a symmetric cross-section), the respective densities ($\rho$) were estimated as 5.6$\times$10$^{9}${\,}cm$^{-3}$ and 4.1$\times$10$^{9}${\,}cm$^{-3}$ by following $\rho = \sqrt{EM/w}$. The width of the loop was calculated from the Full Width at Half Maximum (FWHM) of a Gaussian fitted to the cross-sectional intensity profile from the AIA 94{\,}{\AA} channel near the loop apex. This value did not change much between the two instants and was found to be about 8.5$\pm$0.3 Mm at 17:40 UT and 8.4$\pm$0.3 Mm at 18:20 UT. Using the temperature and density values, along with the oscillation periods derived from the AIA 94{\,}{\AA} channel for the two phases, we solve the dispersion relation for slow waves in the presence of thermal conduction damping \citep[e.g., see Eq. 5 of][]{2014ApJ...789..118K}. Since the oscillations in the present case are standing, we seek solutions for the angular frequency, $\omega$, and then deduce the damping time from the imaginary part of the appropriate root. We use a standard value for $\gamma$ ($=$5/3) in these calculations. Thus obtained theoretical damping times ($\tau^{\prime}$) and other relevant parameters of the oscillation are listed in Table{\,}\ref{tab1}. The derived $\tau^{\prime}$ values are 945{\,}s and 2396{\,}s, respectively, for the two phases of the oscillation. An important aspect to consider here is that the oscillation during the sloshing phase is not a simple sinusoidal wave but rather a travelling impulsive perturbation. Following Fourier's theorem, such perturbation can be expressed as a sum of multiple harmonic oscillations. Since the amplitudes of higher harmonics tend to get lower (see for e.g., \citet{2019ApJ...874L...1N}), we consider the first 10 harmonics to effectively represent the sloshing perturbation. The resultant damping time in such a scenario is estimated to be 882{\,}s which is slightly lower than that obtained from the fundamental period alone. Nevertheless, the theoretical values are on the same order as those obtained from the observations and importantly demonstrate that a factor of two increase in damping time can be readily explained by the decrease in the efficiency of thermal conduction as the plasma cools down. Moreover, additional damping mechanisms such as the compressive viscosity \citep{2018ApJ...860..107W} or thermal misbalance \citep{2017ApJ...849...62N, 2019A&A...628A.133K} can further reduce these damping times possibly closing in the gap with the observations. It may be noted that a direct comparison of these theoretical damping time values with those from observations is not trivial as the dominant emission in a particular wavelength channel could be coming from a different temperature plasma depending on the corresponding filter response function. Additionally, the square root dependence of the sound speed on temperature implies a reduction in the sound speed by a factor of 1.33 from the sloshing to the standing phase, which should effectively increase the oscillation period by the same factor. During this time, the observations reveal an increase in the oscillation period by 1.28 times showing a good agreement with the theory.

As mentioned earlier, a sloshing perturbation can be understood as the simultaneous existence of multiple harmonic oscillations. Under this view, any damping mechanism that effectively dissipates higher harmonics (shorter periods) faster would assist in quickly establishing the fundamental standing mode in a coronal loop. The compressive viscosity is one such mechanism. Using 1D nonlinear MHD simulations \citet{2018ApJ...860..107W} have shown that largely enhanced compressive viscosity could explain their earlier observations \citep{2015ApJ...811L..13W} where a standing oscillation was found to appear immediately after the initial perturbation due to a flare. In the present scenario, however, we see a gradual transformation of the sloshing oscillation into a standing wave. This is perhaps largely in agreement with the model 1 of \citet{2018ApJ...860..107W} where classical values for transport coefficients (thermal conduction and compressive viscosity) are used. Additionally, here, we speculate that the steep decrease in plasma temperature during the initial few cycles has a role in the transformation. It can be shown that in the strong thermal conduction limit, the damping of slow waves due to thermal conduction is independent of the oscillation period whereas the same in the weak thermal conduction limit, is very effective at shorter periods \citep[see Table 1 of][]{2014ApJ...789..118K, 2016ApJ...820...13M}. Although the observed decrease in temperature does not support such extreme changes in thermal conduction, the effect of the cooling plasma on the dissipation of higher harmonics has to be explored in detail through a numerical study. We plan to do this in a follow up work.

\subsection{Phase difference}
We also investigate if there is any phase difference between the oscillations in the 131{\,}{\AA} and the 94{\,}{\AA} channels. For this purpose, the time series at each spatial position from the 131{\,}{\AA} channel is cross correlated against that from the 94{\,}{\AA} channel and the respective time lag corresponding to a peak in the cross correlation are noted. Because the oscillation is observed for a shorter duration in the 131{\,}{\AA} channel, we only consider the section of the time series marked by the orange solid lines in Fig.{\,}\ref{tdmaps}c.
\begin{figure*}
\centering
\includegraphics[scale=0.47]{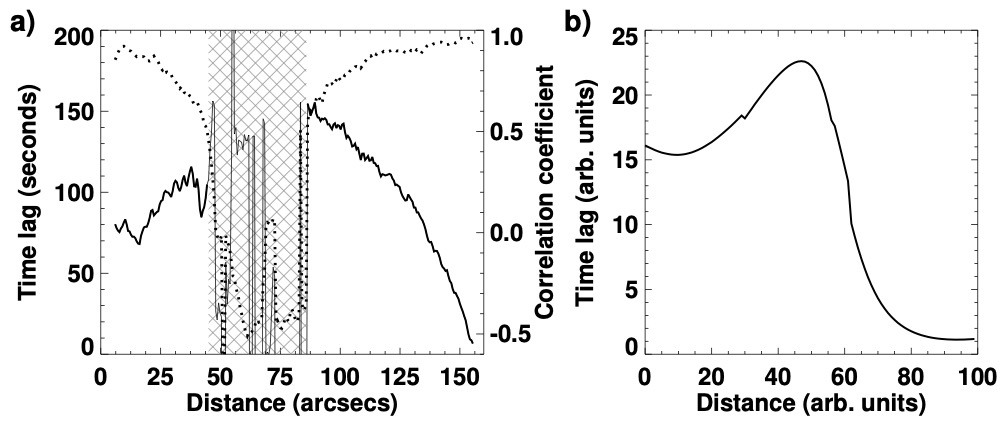} 
\caption{a) Time lag between the oscillation in the 131{\,}{\AA} and the 94{\,}{\AA} channels as a function of distance along the loop. The dotted line shows the corresponding peak cross-correlation coefficient values with their scale on the right. The cross-hatched region highlights a section around the apex where the cross correlation coefficient drops below 0.5. The time lag within this region is less emphasized. b) Time lag obtained from the model (see Eq.{\,}\ref{eq1}) by cross correlating a sloshing oscillation with that propagating at a lower speed.}
\label{phdiff}
\end{figure*}
The obtained time lag and the associated cross-correlation coefficient are plotted in Fig.{\,}\ref{phdiff}a as a function of distance along the loop. As can be seen, the time lag (solid line) is positive and appears to increase from the foot points towards the apex around which it displays sudden random changes. The cross correlation coefficient (dotted line) is positive and close to 1.0 (implying a high correlation) near the foot points but gradually drops to a lower value towards the apex where it also displays sudden random changes. The lower oscillation amplitudes and the possible (de)merging of the oncoming and forward going perturbations near the loop apex are likely the reasons behind the poor correlation near the loop apex. Ignoring this region where the cross correlation coefficient drops below 0.5 (i.e., within the cross-hatched region), the positive time lag values indicate that the oscillation in the 94{\,}{\AA} channel is lagging behind that in the 131{\,}{\AA} channel.

Since the 94{\,}{\AA} channel predominantly captures emission from relatively colder plasma, the perturbation in this channel could be propagating at slower speeds and consequently, positive phase lags with respect to the oscillation in the 131{\,}{\AA} channel are expected. However, the reason for the increase in time lag from the foot points to the apex is not immediately obvious. Also, there is an asymmetry in the time lag between the two foot points with higher values obtained near the left foot point (as in Fig.{\,}\ref{tdmaps}). In order to understand this behaviour, we construct two time-distance maps from our sloshing oscillation model (see Eq.{\,}\ref{eq1}) using two different speeds and perform a cross correlation between them. The obtained time lag as a function of distance is shown in Fig.{\,}\ref{phdiff}b. Although, the distance and the time lag are in arbitrary units, it is evident that the time lag peaks near the centre and the asymmetry in the values between the left and right edges is also  clearly reproduced. We note that this is an average behaviour resulting from the gradually increasing time lag as the perturbation propagates between the two end points. Additionally, we find that this spatial dependence is sensitive to the section of the time series considered, particularly the starting phase and the ending phase of the oscillation. While the time lag values are positive in all cases (indicating the slower perturbation is lagging behind), a minimum time lag near the centre is found in some cases, and the asymmetry between the foot points is flipped with larger lag on the right edge in other cases. Unfortunately, because of the strong damping and low cycle count in observations, we have little freedom to verify this changing shape but nevertheless, we can conclude that the observed time lag and its spatial dependence both indicate lower phase speed in the 94{\,}{\AA} channel. One may perhaps use the actual time lag values to extract more information (e.g., difference in phase speed between the two channels) but, as the plasma properties are varying with time, an advanced model would be necessary.

\subsection{Multi-thermal behaviour}
As mentioned before, there are significant differences in the appearance of the oscillation in the AIA 131{\,}{\AA} and 94{\,}{\AA} channels. The perturbation is barely visible beyond the sloshing phase in the 131{\,}{\AA} channel. During this phase, the oscillation period is approximately the same in both the channels but the damping time is much shorter in the 131{\,}{\AA} channel. The phase difference analysis reveals that the propagation speed is lower in the 94{\,}{\AA} channel. Since the general appearance of the loop is congruent between the channels, this differential behaviour in oscillation properties is indicative of multi-thermal structure within the loop. Indeed, the plasma temperature near the loop apex exhibits a three-part structure during the initial phase of the oscillation (see Fig.{\,}\ref{emte}b) supporting the multi-thermal nature. Yet, the complete absence of standing phase of the oscillation, as if we are observing a distinct loop structure in the 131{\,}{\AA} channel, is difficult to comprehend.

The faster decrease of oscillation amplitude observed in the 131{\,}{\AA} channel definitely supports a quicker disappearance of the oscillation but as we demonstrate in the following that is only part of the reason.
\begin{figure*}
\centering
\includegraphics[scale=0.47]{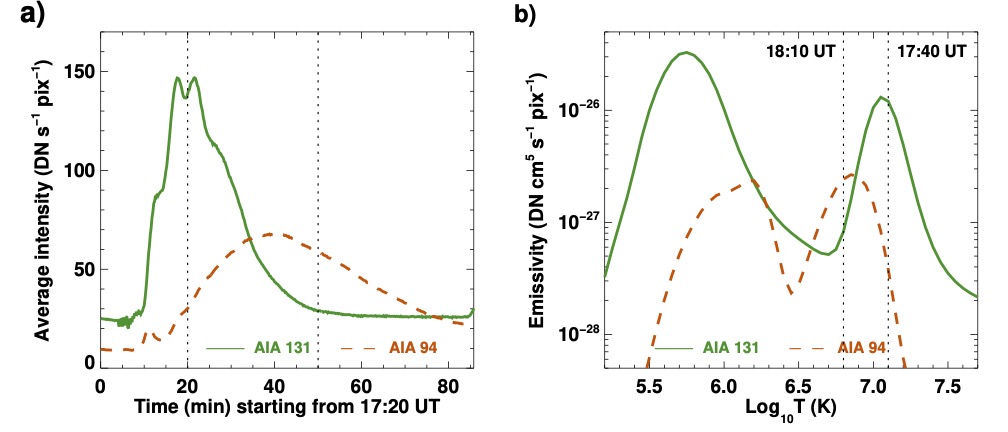} 
\caption{a) Average light curves obtained from a square region near the loop apex marked in Fig.{\,}\ref{emte}. The vertical dotted lines represent two specific instants between which the AIA 131{\,}{\AA} intensity dropped significantly. b) AIA filter response curves for the 131{\,}{\AA} and 94 {\,}{\AA} channels. The vertical dotted lines in this plot represent the plasma temperatures at the two instants marked in panel a. }
\label{iresp}
\end{figure*}
In Fig.{\,}\ref{iresp}a we plot the average light curves for both 131{\,}{\AA} and 94{\,}{\AA} channels obtained from the same box region near the loop apex (see Fig.{\,}\ref{emte}) where the evolution of temperature and emission measure were studied. Evidently, the intensity in AIA 131{\,}{\AA} channel peaks at about 17:40 UT and quickly drops to the pre-flare level by about 18:10 UT. These two instants are marked by vertical dotted lines in the figure. On the other hand, the intensity in AIA 94{\,}{\AA} channel gradually increases, peaks much later than that in the 131{\,}{\AA} channel, and slowly decreases, maintaining above pre-flare level values until the end of our dataset. In Fig.{\,}\ref{iresp}b, we plot the temperature response functions (version 10) of both the AIA channels. These are double peaked with the second peak in the 131{\,}{\AA} channel corresponding to slightly hotter temperatures as compared to that in the 94{\,}{\AA} channel. The two instants marked in Fig.{\,}\ref{iresp}a are also shown in this plot by vertical dotted lines considering the respective dominant plasma temperatures. As can be seen, the expected emissivities vary significantly between the two instants in both the channels. In particular, the decrease in temperature by 18:10 UT has resulted in a significant reduction in the emissivity (by more than an order of magnitude) in the 131{\,}{\AA} channel whereas the same has led to enhanced emissivity in 94{\,}{\AA} channel. As a result, if one considers the corresponding decrease in density (emission measure; see Fig.{\,}\ref{emte}), the reduction in the emission seen by the 131{\,}{\AA} channel is much more pronounced whereas the same in the 94{\,}{\AA} channel is somewhat compensated, thus, explaining the sharper decline in intensity observed in the former channel. Because the 131{\,}{\AA} channel observes hotter plasma, faster damping (shorter damping time) is expected due to thermal conduction. The low intensity further compounded by the lower oscillation amplitude made it difficult for the oscillation to be detected in the 131{\,}{\AA} channel post 18:10 UT. We believe, had the intensity remained sufficiently large, the standing phase of the oscillation would have been visible in the 131{\,}{\AA} channel as well. So the stark contrast in the appearance of oscillation in the two AIA channels is due to an interplay between the oscillation amplitude, plasma temperature, and the filter response curves. Therefore, one should be careful while interpreting the observations of sloshing oscillations that do not appear to transform into a standing wave.

\section{Conclusions}
\label{concl}
We study the compressive plasma oscillation in a hot coronal loop triggered by a C-class flare near one of its foot points. Recent studies using high-resolution imaging observations by SDO/AIA have resulted in interpretations of such oscillations either in terms of a reflected propagating (sloshing) wave \citep{2013ApJ...779L...7K} or a standing wave \citep{2015ApJ...811L..13W}. The sloshing oscillations are expected to eventually transform into a standing wave \citep{2018ApJ...860..107W} but the observational evidence for this has not been very clear until now. It has also been argued that the sloshing oscillations could be entirely a different class of oscillations \citep{2019ApJ...874L...1N}. Here, we revisit the event studied by \citet{2013ApJ...779L...7K}, extend their dataset to a longer duration, and analyse the oscillation properties in both the AIA 131{\,}{\AA} and 94{\,}{\AA} channels in detail. Our results indicate a number of interesting new properties of the oscillation which was earlier interpreted as a reflected propagating wave by \citet{2013ApJ...779L...7K}. The main conclusions of our study are listed below.
\begin{enumerate}
\item We find that the oscillation is visible up to six cycles in the AIA 94{\,}{\AA} channel in contrast to only three cycles observed in the AIA 131{\,}{\AA} channel as reported earlier. The longer duration dataset used here has allowed us to reveal this observation. Furthermore, the oscillation evidently transforms into a standing wave during the later part of the time series in the AIA 94{\,}{\AA} channel. This suggests that the sloshing oscillations are perhaps not an independent class of oscillations.
\item As we observe the oscillation for a longer duration, we are able to extract the oscillation properties separately for the sloshing phase and the standing phase. We find that the oscillation period and the damping time are longer during the latter phase. 
\item For the sloshing phase, we demonstrate that a damping sinusoid function fitted to the time series at any particular spatial location does not \textit{always} provide an accurate description of the oscillation properties. Instead, we provide an analytical expression (Equation{\,}\ref{eq2}) that may be generally used to fit the time series near the loop foot point in order to derive the oscillation properties.
\item Using DEM analysis, we find that the plasma is cooling down during the oscillation. Considering this the observed oscillation properties and the associated changes between the sloshing phase and the standing phase are shown to be compatible with damping due to thermal conduction.
\item The phase difference between the oscillation in the 131{\,}{\AA} and the 94{\,}{\AA} channels indicate lower propagation speed in the latter channel. The distinct oscillation characteristics observed across the two AIA channels imply a multi-thermal nature of the loop. The standing phase of the oscillation is not detectable in the 131{\,}{\AA} channel due to a steep decline in the loop intensity (caused by the cooling plasma) in addition to the faster decay in the oscillation amplitude. 
\end{enumerate}
 
\acknowledgements 
The authors thank the anonymous referee for their useful comments. SKP thanks R. Keppens for helpful discussions. SKP is grateful to FWO Vlaanderen for a senior postdoctoral fellowship (No. 12ZF420N). TVD was supported by the European Research Council (ERC) under the European Union's Horizon 2020 research and innovation programme (grant agreement No 724326) and the C1 grant TRACEspace of Internal Funds KU Leuven. AIA data used here are courtesy of NASA/SDO and the AIA science team. We acknowledge the use of pipeline developed by Rob Rutten to extract, process, and co-align AIA cutout data.

\end{document}